# High-Pressure Synthesis of a Massive and Non-Symmorphic Dirac Semimetal Candidate MoP$_4$

Alex Hiro Mayo[1,2], Jon Alexander Richards[1], Hidefumi Takahashi[1], and Shintaro Ishiwata[1]

[1]*Division of Materials Physics and Center for Spintronics Research Network (CSRN), Graduate School of Engineering Science, Osaka University, Osaka 560-8531, Japan*
[2]*Department of Applied Physics, The University of Tokyo, Tokyo 113-8656, Japan*

Single crystal and polycrystalline samples of MoP$_4$ with a black-phosphorus-derived structure have been successfully synthesized by a high-pressure technique. The polycrystalline samples show a large positive magnetoresistance and a small negative Seebeck coefficient at low temperatures, reflecting a semi-metallic nature with high-mobility electrons. Consistent with the transport properties, the band structure calculation reveals a semi-metallic state with the presence of two types of Dirac nodes slightly below the Fermi level. The Dirac node along the Γ-X direction normal to the phosphorus layers is gapped out in the presence of spin-orbit coupling (SOC), whereas the band crossing at the Z-point is immune to SOC because of the non-symmorphic symmetry. This work demonstrates a great potential of phosphorus-based layered Zintl compounds for topological semimetal candidates allowing chemical band engineering.

In the field of condensed matter physics, the search for materials that exhibit novel quantum transport phenomena has been vigorously pursued. Among them, topological semimetals (TSMs), such as Dirac, Weyl, and nodal-line semimetals, are expected to have the potential to realize dramatically higher performing electronic devices owing to the unique transport properties associated with band topology.[1,2] The band dispersion of TSMs differs from that of ordinary materials, in that it exhibits a linear dispersion near the Fermi level, hosting relativistic quasiparticles. One of the promising platforms for realizing such an electronic structure is honeycomb-based materials, most well represented by graphene and phosphorene related materials.[3,4]

From the above perspective, we focused on black phosphorus (BP), a layered material that is predicted to become a Dirac semimetal under high pressures.[5–7] BP is a bulk material consisting of stacked phosphorene atomic layers. The interest of BP can be seen not only in the elemental material[8,9] but also in its related compounds[10–12]: in Zintl compounds, anions are covalently bonded to form clusters that satisfy the octet rule and form crystals by ionic bonding with cations.[13] In the case of phosphorus-based Zintl compounds, the anionic clusters have structures derived from the flexible bond angles of phosphorus, which allows a great variety of structures to be formed. Furthermore, such an approach enables the exploration of novel functionalities by combining various metal ions located between the phosphorus layers without destroying the Dirac-like nature derived from the closed-shell characteristic of the layers. Recently, magnetic and non-magnetic TSMs based on BP-related structures have been reported,[14–16] providing rich opportunities for the investigation of TSMs. Also, a series of transition metal tetraphosphides $TM$P$_4$ ($TM$ = V, Cr, Mn, Fe, Mo) has been reported in previous works,[17–20] proving the potential of phosphorus-based materials providing a wide variety of crystal structures. However, while some calculations predict that such systems are topological nodal-line semimetal candidates,[21,22] detailed physical properties remain elusive and have not been investigated. In this study, we report the synthesis, transport properties, and electronic structure of a phosphorus-based Dirac semimetal candidate MoP$_4$. The material demonstrates a large non-quadratic positive magnetoresistance and a sign change in the temperature dependence of the Seebeck coefficient, indicating that the system is a semimetal with coexisting electrons and holes. These interpretations were verified by band calculations, also suggesting the presence of symmetry-protected Dirac points near the Fermi level.

Single crystal and polycrystalline samples of MoP$_4$ were prepared by the high-pressure synthesis method.[17] The synthesis procedures implemented in this work are as follows. A mixture of Mo powder (99.9 % purity) and red phosphorus pieces (99.9999 % purity) in a stoichiometric molar ratio of 1 : 4 was weighed and put into a pyrophyllite crucible. The crucible was then placed in a graphite tube heater. The sample assembly was set in a pyrophyllite cube of 12.7 × 12.7 × 12.7 mm$^3$ and high pressure was applied using a cubic multi-anvil apparatus. Under an applied pressure of 4 GPa, the sample was heated to 900 °C within 2 minutes. For single crystal samples, the temperature was kept at 900 °C for 1 hour and was then quenched to room temperature before releasing the pressure. For polycrystalline samples, after the temperature was kept for 30 minutes, the sample was cooled down to 700 °C at a rate of 100 °C/h, followed by a quenching process to room temperature before releasing the pressure. The structural parameters of the obtained single crystal were refined at room temperature using a Rigaku XtaLAB-mini II diffractometer with graphite monochromated Mo $K\alpha$ radiation. The polycrystalline sample was confirmed to be a single phase by powder x-ray diffraction (XRD) method using a Bruker D8 advance diffractometer with Cu $K\alpha$ radiation at room temperature. Physical properties were measured using the polycrystalline sample. The transport properties were measured using a Magnetic Property Measurement System Electrical Transport Option (Quantum Design). The longitudinal and Hall resistivities were measured by the standard five-probe method. The Seebeck coefficient was measured by the steady-state method. The relativistic and non-relativistic bulk electronic structure calculations were performed within the density functional theory (DFT) using the Perdew-Burke-Ernzerhof (PBE) exchange-correlation functional as implemented in the Quantum ESPRESSO code.[23–25] The projector augmented wave (PAW) method has been used to account for the treatment of core electrons.[26] The cut-off energy for plane waves forming the basis set was set to 50 Ry. The lattice parameters and atomic positions were taken from the single crystal XRD experiment. The Brillouin zone (BZ) was sampled by an 8 × 8 × 8 $k$-mesh for the PWscf calculation and a 16 × 16 × 16 $k$-mesh for the Fermi surface visualization by the FermiSurfer package.[27] The crystal structures were visualized using VESTA.[28]

Figures 1(a) and (b) show the crystal structure based on the parameters obtained from the single crystal XRD experiment, using the sample shown in the inset of panel (d). The data were well refined by the monoclinic space-group type $C2/c$, giving rise to structure parameters $a$ = 5.3131(6) Å, $b$ = 11.1588(9) Å, $c$ = 5.8343(6) Å, and $\beta$ = 110.638(12)°, consistent with previous reports.[17,29] Details of the structure refinements are given in Tables I and II. Fig. 1(d) shows the powder XRD profile for the polycrystalline sample. All of the diffraction peaks were assigned by the Le Bail refinement, revealing that the sample is essentially a single phase. As shown in Fig. 1(a), the unit cell contains double phosphorus layers sandwiching the Mo atoms, which can be correlated with each other by the $c$-glide plane of ($x$, 0, $z$) normal to the $b$-axis. The phosphorus layers are highly buckled as those in BP. On the other hand, unlike BP with each phosphorus atom having three covalent bonds, MoP$_4$ has two types of phosphorus atoms, P(1) and P(2), where P(1) is covalently bonded to three P atoms along with a dative bond to one Mo atom through its single lone pair, while P(2) is covalently bonded to two P atoms along with dative bonds to two Mo atoms through its two lone pairs (Figs. 1 (b) and (c)). These features can be viewed as a characteristic of Zintl compounds satisfying the octet rule for the anions. By assigning the formal valence of +2 to the Mo ion, alike Mg$^{2+}$ in the related material MgP$_4$,[30] it is reasonable to presume that the valence electrons in P(1) and P(2) form $sp^3$-like orbitals with single and double lone pairs per atom, respectively; i.e., the nominal valence of P(1) is zero and that of P(2) is -1.[17,29]

Transport properties have been measured using the obtained polycrystalline sample. Figures 2(a) and (b) show the temperature dependences of the resistivity $\rho_{xx}$ and the Seebeck coefficient $S$, respectively. $\rho_{xx}$ shows metallic behavior with a residual resistivity ratio of $\rho_{xx}(T$ = 300 K$)/\rho_{xx}$(1.8 K) = 3.5. Under the application of a magnetic field perpendicular to the current direction, $\rho_{xx}$ shows a large positive magnetoresistance (MR) of $\Delta\rho/\rho_0 \equiv \{\rho_{xx}(B = 7 \text{ T}) - \rho_{xx}(0 \text{ T})\}/\rho_{xx}(0 \text{ T}) = 158$ % at 1.8 K. The Seebeck coefficient is as small as a few µV/K throughout the measured temperature region and shows a sign change at around 130 K, turning negative at lower temperatures. This behavior indicates that MoP$_4$ is an electron-hole system, possibly possessing high-mobility electron pockets contributing to the electronic transport at low temperatures.

In order to obtain further information on the band structure near the Fermi level, we have conducted magneto-transport measurements and two-carrier analyses. Figures 3(a) and (b) show the magnetic field dependences of longitudinal ($\rho_{xx}$) and Hall ($\rho_{yx}$) resistivities, respectively, measured at 1.8 K. $\rho_{xx}$ shows non-saturating behavior up to $B = 7$ T, with a $B$ dependence of $\rho_{xx} = \rho_0 + a \cdot B^{1.53}$, where $\rho_0$ and $a$ are the resistivity at zero field and a coefficient, respectively. In the conventional two-carrier model considering electrons and holes, the MR can be represented by the following equation[31,32]:

$$\frac{\Delta\rho}{\rho_0} = \frac{B^2 n_e n_h |\mu_e||\mu_h|(|\mu_e| + |\mu_h|)^2}{(|\mu_e|n_e + |\mu_h|n_h)^2 + B^2 \mu_e^2 \mu_h^2 (n_e - n_h)^2},$$

where $n_e$ ($n_h$) and $\mu_e$ ($\mu_h$) are the charge carrier concentration and the mobility of electrons (holes), respectively. In the case of an ideally compensated semimetal ($n_e = n_h$), the MR expression is reduced to a quadratic $B$ dependence, whereas incomplete compensation ($n_e \neq n_h$) leads to an exponent smaller than 2, deviating from the quadratic behavior. Given that the exponent of $B$ is 1.53 < 2, the latter condition is probably the case for the measured sample. For $\rho_{yx}$, we have carried out a two-carrier model analysis based on the scheme presented in the paper by G. Eguchi and S. Paschen.[33] In this scheme, $N \equiv (n_1 - n_2)/(n_1 + n_2)$, $M \equiv (\mu_1 - \mu_2)/(\mu_1 + \mu_2)$, $R_H \equiv \lim_{B\to 0} \rho_{yx}/B$, and $\mu_H \equiv \lim_{B\to 0} \rho_{yx}/(\rho_{xx}B)$ are applied as the fitting parameters instead of $n_e$, $n_h$, $\mu_e$, and $\mu_h$. Here, $n_1$ ($n_2$) and $\mu_1$ ($\mu_2$) are the charge carrier concentration and mobility of the majority (minority) carriers, respectively. Since $R_H$ and $\mu_H$ can be fixed to the values directly obtained from experimental data, the free parameters are reduced to $N$ and $M$. Using these parameters, $\rho_{yx}(B)$ can be formulated as[33]

$$\rho_{yx}(B) = \frac{2}{n_+ q_1} \left[ \frac{(1+N)(1+M)^2 + (q_2/q_1)(1-N)(1-M)^2 + [(1+N) + (q_2/q_1)(1-N)](1-M^2)^2(\mu_+ B)^2}{[(1+N)(1+M) + (q_2/q_1)(1-N)(1-M)]^2 + [(1+N) + (q_2/q_1)(1-N)]^2(1-M^2)^2(\mu_+ B)^2} \right] B,$$

with defining $n_+$ and $\mu_+$ as

$$n_+ \equiv n_1 + n_2 = \frac{2}{q_1 R_H} \left[ \frac{(1+N)(1+M)^2 + (q_2/q_1)(1-N)(1-M)^2}{[(1+N)(1+M) + (q_2/q_1)(1-N)(1-M)]^2} \right],$$

and

$$\mu_+ \equiv \frac{\mu_1 + \mu_2}{2} = \mu_H \frac{(1+N)(1+M) + (q_2/q_1)(1-N)(1-M)}{(1+N)(1+M)^2 + (q_2/q_1)(1-N)(1-M)^2}.$$

$q_1$, $q_2 = \pm e$ are charge carrier types, where $e$ is the elementary charge. Based on this formulation, we obtain $n_1 = n_h = 2.78 \times 10^{20}$ cm$^{-3}$, $n_2 = n_e = 9.56 \times 10^{19}$ cm$^{-3}$, $\mu_1 = \mu_h = 88.6$ cm$^2$/Vs, $\mu_2 = \mu_e$ = -673 cm$^2$/Vs, which appears to be reasonably consistent with the observed sign change behavior in the Seebeck measurement.

Such behavior in the transport measurements can be associated with a semi-metallic nature of the material, as expected from the band calculation shown in Figure 4. The electronic structure clearly demonstrates that MoP$_4$ is a semimetal with small electron and hole pockets at the Fermi level (see Figs. 4(a) and (c)). One of the key features is the occurrence of a band inversion around the Γ-point. As a result, a gapless Dirac dispersion appears slightly below $E_F$ along the Γ-X direction, which corresponds to the inter-layer direction parallel to the $b$-axis, when spin-orbit coupling (SOC) is not taken into account (red dashed line in Figs. 4(a) and (d)). When SOC is included (black solid line), the band crossing gaps out, resulting in a massive Dirac dispersion. Another key feature can be seen at the Z-point on the BZ boundary, where Γ-Z is the intra-layer direction parallel to the $c^*$-axis. Here, we can see the presence of another Dirac-like feature below $E_F$. In this case, however, such feature is caused by a folding of the BZ enforced by non-symmorphic symmetry,[34–37] rather than an inversion of valence and conduction bands. Therefore, the Dirac node at the Z-point is immune to SOC.

The observations of the high electron mobility estimated from the magneto-transport measurements and the negative Seebeck coefficient at low temperatures are presumably attributable to the highly dispersive bands stemming from the Dirac-like electron pocket along Γ-X, dominating the electronic transport compared to the contribution from the relatively low mobility hole pocket. As seen in the calculated partial density of states shown in Fig. 4(b), the electronic states near the Fermi level are dominated by the Mo-$d$ and P-$p$ orbitals. Thus, it can be likely said that the Mo-P bonds bridging the phosphorus layers play an important role on the characteristic band structure in this material.

In conclusion, we have successfully obtained single crystals and polycrystals of a Dirac semimetal candidate MoP$_4$ by means of high-pressure synthesis. The single crystal x-ray refinement agrees well with the previous report, and polycrystalline samples were proven to be a single phase by powder XRD. The non-quadratic field dependence of the resistivity and the temperature dependent sign change in the Seebeck coefficient were successively characterized by the two-carrier model, which indicates the predominance of high-mobility electrons at the lowest temperature. Furthermore, this model was supported by the first-principles calculations indicating the semi-metallic band structure yielding the massive and the symmetry-protected Dirac electrons perpendicular and parallel to the phosphorus layers, respectively. This work demonstrates that phosphorus-based layered Zintl compounds are a fertile platform for exploring novel topological semimetals, owing to their closed-shell electronic nature and the structural variety derived from phosphorus covalent bonds involving a wide variety of transition metal ions.[17–19]

**Acknowledgement**

This study was supported in part by KAKENHI (Grant No. 17H06137, 19H02424, 19K14652, 20K03802 and 21H01030), the Research Foundation for the Electrotechnology of Chubu, the Murata Science Foundation, and the Asahi Glass Foundation.

*E-mail: mayo@qm.mp.es.osaka-u.ac.jp, ishiwata@mp.es.osaka-u.ac.jp

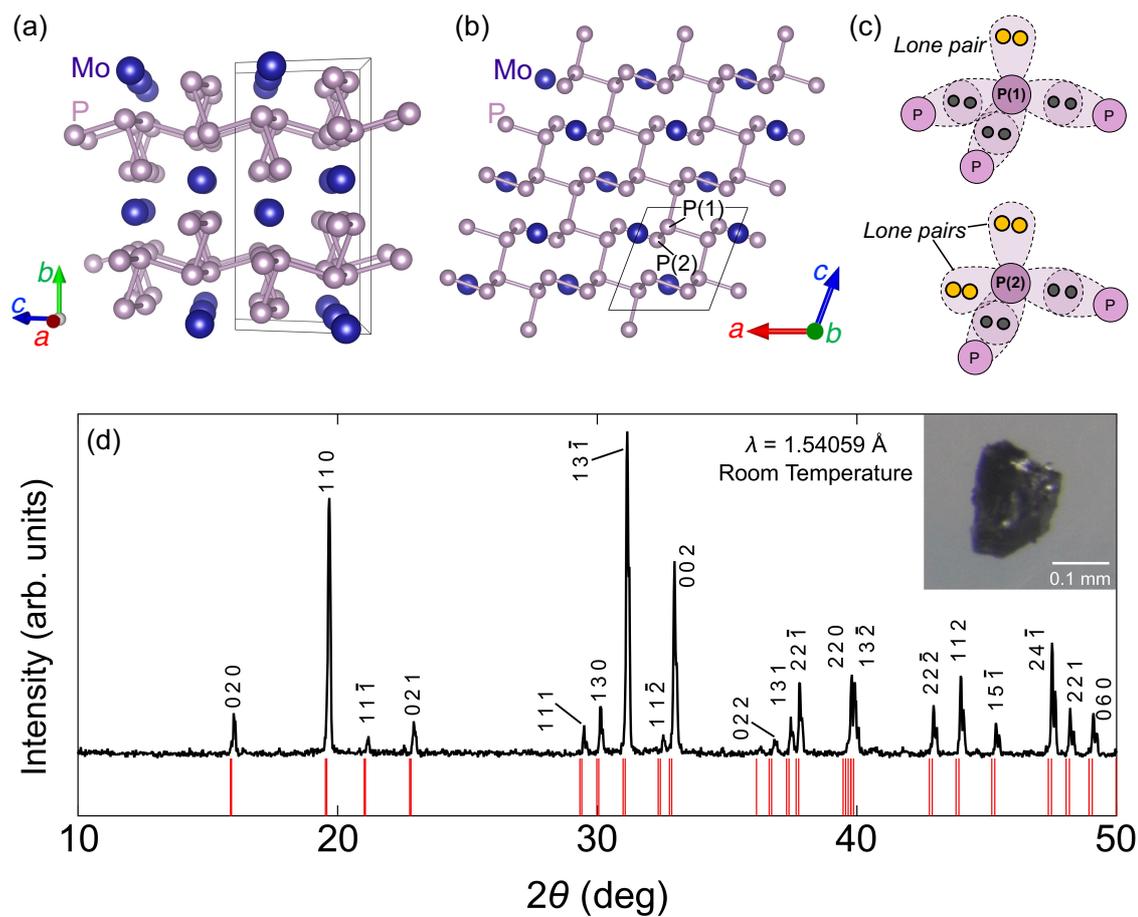

Fig. 1. (Color online) (a) Side view and (b) top view of the layered crystal structure of MoP$_4$ obtained from single crystal x-ray diffraction. The two inequivalent phosphorus sites P(1) and P(2) are denoted in panel (b). (c) Schematic illustrations of the bonding characters of P(1) (upper panel) and P(2) (lower panel). (d) Powder x-ray diffraction profile of MoP$_4$. Red ticks correspond to the calculated peak positions. A picture of a single crystal sample is given in the inset.

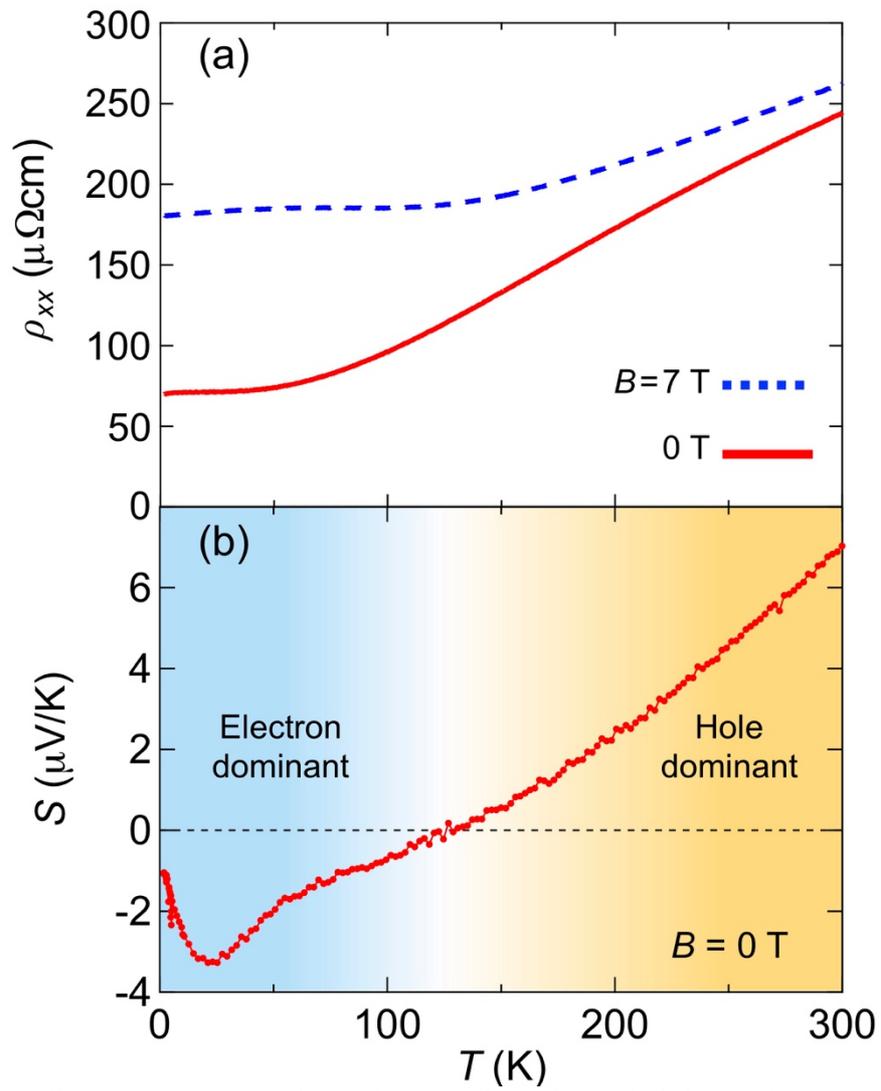

Fig. 2. (Color online) Temperature dependences of (a) the resistivity under $B = 0$ and 7 T and (b) the Seebeck coefficient.

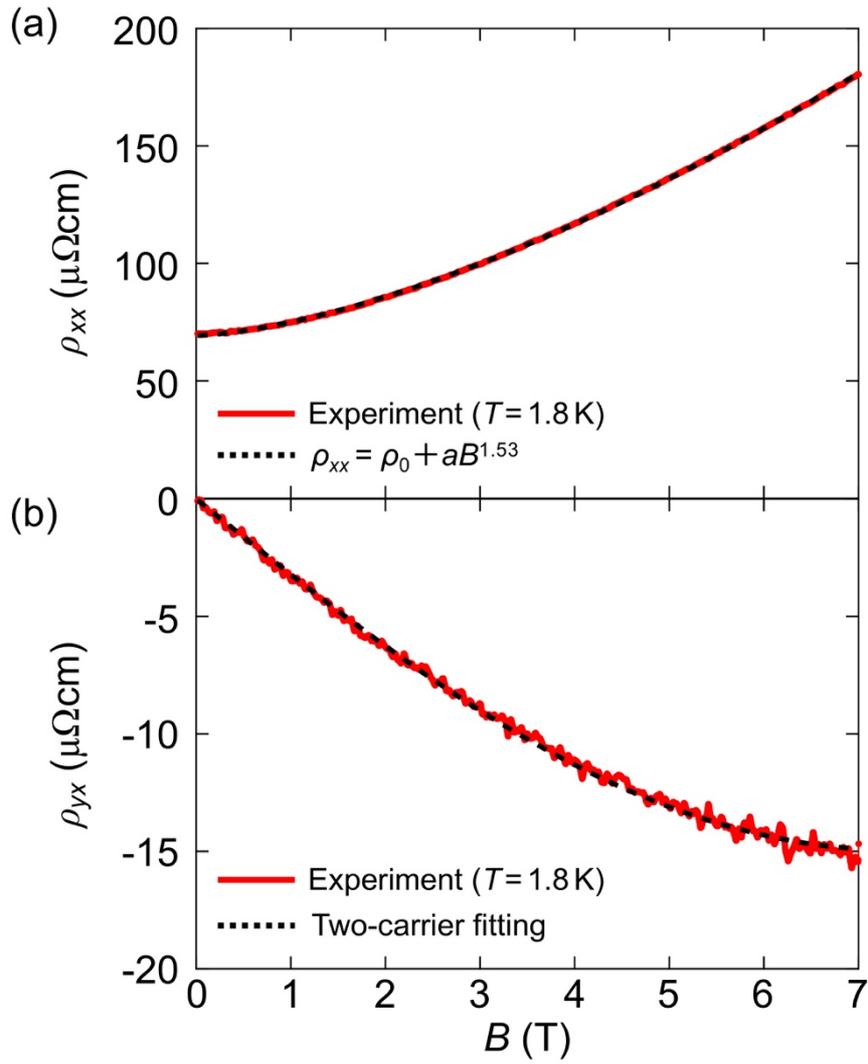

Fig. 3. (Color online) Magnetic field dependences of (a) longitudinal and (b) Hall resistivities, measured at $T$ = 1.8 K.

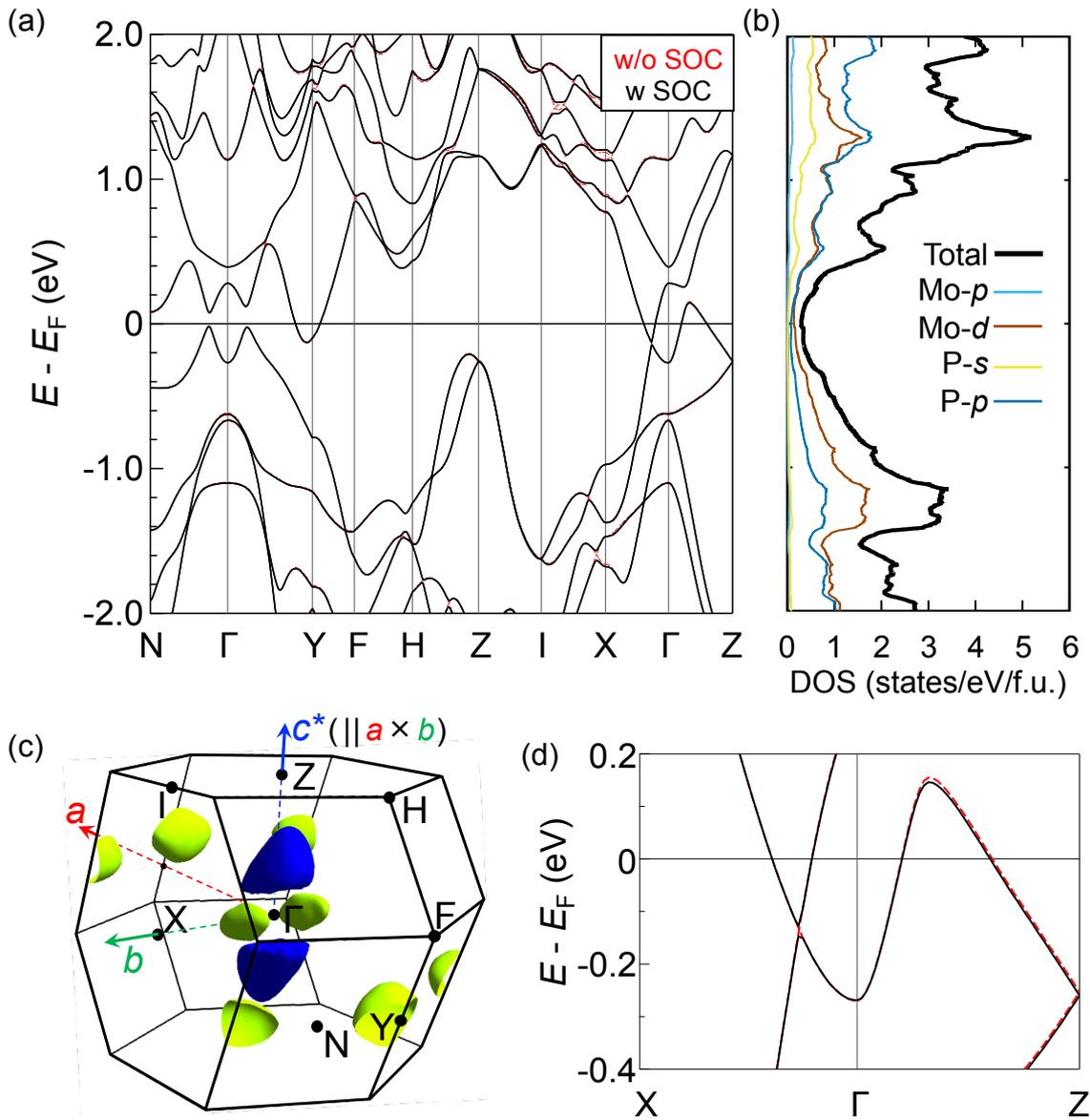

Fig. 4. (Color online) (a) Band structure of MoP$_4$ calculated with (black solid line) and without (red dashed line) SOC. (b) Total and partial density of states (DOS) for MoP$_4$ with SOC. (c) Fermi surface of MoP$_4$ in its monoclinic Brillouin zone.[38)] The yellow (blue) surfaces correspond to electron (hole) pockets. (d) Magnified view of the bands near the Fermi level along X-Γ-Z.

Table I. Crystallographic data and structure refinement for MoP$_4$ at room temperature.

| Formula | MoP$_4$ |
|---|---|
| Crystal system | Monoclinic |
| Space-group type | $C2/c$ (No. 15) |
| $a$ (Å) | 5.3131(6) |
| $b$ (Å) | 11.1588(9) |
| $c$ (Å) | 5.8343(6) |
| $\alpha$ | 90° |
| $\beta$ | 110.638(12)° |
| $\gamma$ | 90° |
| $V$ (Å$^3$), $Z$ | 323.70(6), 4 |
| $d_{calc}$ (g/cm$^3$) | 4.511 |
| $\mu$ (mm$^{-1}$) | 5.734 |
| $\lambda$ Mo $K\alpha$ (Å) | 0.71073 |
| $T$ (K) | 293(2) |
| Theta range for data collection | 3.652 to 29.862° |
| Independent reflections | 452 |
| Data / restraints / parameters | 452 / 0 / 24 |
| GOF on $F^2$ | 1.245 |
| $R_1$/w$R_2$ [$I \geq 2\sigma(I)$] | 0.0502/0.1199 |
| $R_1$/w$R_2$ (all data) | 0.0513/0.1210 |

Table II. Atomic coordinates and equivalent isotropic displacement parameter $U_{eq}$ (Å$^2$) for MoP$_4$.

| Atoms | Multiplicity | Wyckoff letter | x | y | z | $U_{eq}$ |
|---|---|---|---|---|---|---|
| Mo | 4 | e | 0 | 0.05877(4) | 1/4 | 0.0005(3) |
| P(1) | 8 | f | 0.2779(2) | 0.22215(11) | 0.1893(2) | 0.0023(3) |
| P(2) | 8 | f | 0.2202(2) | 0.40548(12) | 0.3163(2) | 0.0023(3) |